\documentstyle[12pt]{article}
\begin{document}
 \newcommand{\be}[1]{\begin{equation}\label{#1}}
 \newcommand{\ee}{\end{equation}}
 \newcommand{\beqn}[1]{\begin{eqnarray}\label{#1}}
 \newcommand{\eeqn}{\end{eqnarray}}
\newcommand{\bd}{\begin{displaymath}}
\newcommand{\ed}{\end{displaymath}}
\newcommand{\mat}[4]{\left(\begin{array}{cc}{#1}&{#2}\\{#3}&{#4}\end{array}
\right)}
\newcommand{\matr}[9]{\left(\begin{array}{ccc}{#1}&{#2}&{#3}\\{#4}&{#5}&{#6}\\
{#7}&{#8}&{#9}\end{array}\right)}
 \newcommand{\eps}{\varepsilon}
\newcommand{\ov}{\overline}
\renewcommand{\to}{\rightarrow}
\renewcommand{\thefootnote}{\fnsymbol{footnote}}
%
%
\makeatletter
\newcounter{alphaequation}[equation]
\def\thealphaequation{\theequation\alph{alphaequation}}
%
\def\eqnsystem#1{
\def\@eqnnum{{\rm (\thealphaequation)}}
\def\@@eqncr{\let\@tempa\relax
\ifcase\@eqcnt \def\@tempa{& & &}
\or \def\@tempa{& &}\or \def\@tempa{&}\fi\@tempa
\if@eqnsw\@eqnnum\refstepcounter{alphaequation}\fi
\global\@eqnswtrue\global\@eqcnt=0\cr}
\refstepcounter{equation}
\let\@currentlabel\theequation
\def\@tempb{#1}
\ifx\@tempb\empty\else\label{#1}\fi
\refstepcounter{alphaequation}
\let\@currentlabel\thealphaequation
\global\@eqnswtrue\global\@eqcnt=0
\tabskip\@centering\let\\=\@eqncr
$$\halign to \displaywidth\bgroup
  \@eqnsel\hskip\@centering
  $\displaystyle\tabskip\z@{##}$&\global\@eqcnt\@ne
  \hskip2\arraycolsep\hfil${##}$\hfil&
  \global\@eqcnt\tw@\hskip2\arraycolsep
  $\displaystyle\tabskip\z@{##}$\hfil
  \tabskip\@centering&\llap{##}\tabskip\z@\cr}

\def\endeqnsystem{\@@eqncr\egroup$$\global\@ignoretrue}
\makeatother


\begin{flushright}
INFN-FE 14-94 \\
hep-ph/9412372 \\
November 1994
\end{flushright}
\vspace{15mm}

\begin{center}
{\Large \bf Solving SUSY GUT Problems: } \\
\vspace{4mm}
{\Large \bf Gauge Hierarchy and Fermion Masses} \footnote{
On the basis of talks given at the Int. Workshop
"Physics from Planck Scale to Electroweak Scale",
Warsaw, Poland, 21-24 September 1994, and at the III Trieste Conference
"Recent Developments in the Phenomenology of Particle Physics", Trieste,
Italy, 3-7 October 1994.} \\
\end{center}

\vspace{0.6cm}
\centerline{\large Zurab  Berezhiani }
\vspace{5mm}
\centerline{\tenit Istituto Nazionale di Fisica Nucleare, Sezione Ferrara }
\baselineskip=12pt
\centerline{\tenit 44100 Ferrara, Italy }
\vspace{0.1cm}
\centerline{\tenrm and}
\vspace{0.1cm}
\baselineskip=13pt
\centerline{\tenit Institute of Physics, Georgian Academy of Sciences}
\baselineskip=12pt
\centerline{\tenit 38077 Tbilisi, Georgia  }

\vspace{1.0cm}
\begin{abstract}
The supersymmetric $SU(6)$ model accompanied by the
flavour-blind discrete symmetry $Z_3$ can succesfully deal
with such key problems of SUSY GUTs, as are the
gauge hierarchy/doublet-triplet splitting, $\mu$-problem and
flavour problem. The Higgs doublets arise as Goldstone modes
of the spontaneously broken
{\em accidental} global $SU(6)\times U(6)$ symmetry of the
Higgs superpotential. Their couplings to fermions have
peculiarities leading to the consistent picture of the quark and
lepton masses and mixing, without invoking any of
horizontal symmetry/zero texture concepts. In particular,
the only particle that has direct Yukawa coupling with the
Higgs doublet is top quark. Other fermion
masses appear from the higher order operators, with natural
mass hierarchy. Specific mass formulas are also obtained.
\end{abstract}

\vfil
\rm\baselineskip=14pt






\renewcommand{\thefootnote}{\arabic{footnote})}
\setcounter{footnote}{0}

\newpage

{\large \bf 1. Introduction }
\vspace{0.7cm}

The experimental developments of the last few years have confronted
us with
the impressive phenomenon of gauge coupling crossing \cite{crossing}
in the framework of minimal supersymmetric standard model (MSSM):
at the scale $M_G\simeq 10^{16}\,$GeV, $SU(3)\times SU(2)\times U(1)$
can be consistently embedded in $SU(5)$, which in itself can be
further extended to some larger group at some larger scale. At this
point two ideas, elegant GUT and beautiful SUSY, already long time
wanted each other, succesfully meet. It has become completely
clear that GUT without SUSY is not viable \cite{crossing}:
non-supersymmetric $SU(5)$ is excluded while the larger GUTs require
some intermediate scale as an extra parameter, whereas the SUSY
$SU(5)$ prediction \cite{DRW} for $\sin^2\theta_W$ well agrees
with experiment. On the other hand, SUSY without GUT
(i.e. MSSM directly from the string unification), gives too small
$\sin^2\theta_W$. In this view, it is more attractive to think that
at compactification scale $M\sim 10^{18}\,$GeV, the string theory
first reduces to some SUSY GUT containing $SU(5)$, which then
breaks down to standard $SU(3)\times SU(2)\times U(1)$
at $M_G\simeq 10^{16}\,$GeV.

Besides this experimental hint, grand unification needs
supersymmetry for conceptual reasons \cite{Maiani}, related to
so-called gauge hierarchy problem.
At the level of standard model this is essentially a problem of
the Higgs mass ($\sim M_W$) stability
against radiative corrections (quadratic divergences).
It is removed as soon as one appeals to SUSY, which links the
masses of scalars with those of their fermion superpartners while
the latter are protected by the chiral symmetry.
However, in the context of grand unification the gauge hierarchy
problem rather concerns the origin of scales:
why the electroweak scale is so small as compared to the GUT scale
$M_G$, which in itself is not far from the Planck scale.
This question is inevitably connected with the puzzle of the
so-called doublet-triplet (DT) splitting:
in GUT supermultiplets the MSSM Higgs doublets
are accompanied by the colour triplet partners, which mediate an
unacceptably fast proton decay (especially via $d=5$
operators \cite{dim5}) unless their masses are very
large ($\sim M_G$). In addition light triplets, even being
decoupled from quarks and leptons by some means \cite{Dvali},
would spoil the unification of gauge couplings.

For example, in minimal SUSY $SU(5)$ with fermions $(\ov{5}+10)_i$,
where $i=1,2,3$ is a family index, and Higgs sector consisting
of superfields $\Phi(24)$ and $\phi(5)+\bar{\phi}(\bar{5})$,
the $SU(3)\times SU(2)\times U(1)$ decomposition of the latter
is $\phi(5)=T(3,1)+h_2(1,2)$ and
$\bar{\phi}(\bar{5})=\bar{T}(\bar{3},1)+h_1(1,\bar{2})$, where
$h_{1,2}$ are the MSSM higgs doublets and $T,\bar{T}$ are their triplet
partners. The only source of DT splitting can be the interaction
of $\phi,\bar{\phi}$ with the 24-plet $\Phi$. Indeed, the most
general superpotential of these fields has the form:
\be{SU5}
W_{Higgs}=M \Phi^2 + m\bar{\phi}\phi + \lambda \Phi^3 +
f \bar{\phi}\Phi\phi
\ee
The $SU(5)$ breaking to $SU(3)\times SU(2)\times U(1)$ is given
by supersymmetric vacuum
$\langle \Phi \rangle = V_G \mbox{diag} (2,2,2,-3,-3)$,
$\langle \phi \rangle, \langle \bar{\phi} \rangle = 0$,
with $V_G=2M/3\lambda$.
Then the masses of the $T$ and $h$ fragments are respectively
$m+2fV_G$ and $m-3fV_G$, so that massless doublets require the
relation
\be{fine_tuning}
3\lambda m= 2f M
\ee
and mass of triplet is unavoidably $O(M)$ in this case.
Non-renormalization theorem guarantees that in the exact
supersymmetry limit this constraint is stable against radiative
corrections. However, this so-called technical
solution \cite{Maiani} is nothing but {\em fine tuning} of
different unrelated parameters in the superpotential.

The actual task is to achieve the DT-splitting in a natural way
(without fine tuning) due to symmetry reasons, which implies a
certain choice of GUT symmetry and its field content. Several
attempts have been done, but non of them seems satisfactory.
The sliding singlet scenario \cite{slid} was shown to be
unstable under radiative corrections \cite{noslid}.
The group-theoretical trick known as a missing partner
mechanism \cite{miss} needs rather complicated Higgs sector when
implemented in $SU(5)$. In a most economic way it works in the
{\em flipped} $SU(5)$ \cite{flip}, which is not however a
{\em genuine} GUT unifying the gauge couplings.
The missing VEV mechanism \cite{DW} implemented in SUSY $SO(10)$ also
requires rather artificial Higgs sector if one attempts \cite{Barr}
to support it by some symmetry. In addition,
the 'missing' (partner, VEV) mechanisms once being motivated by
symmetry reasons so that the Higgs doublets
are strictly massless in the exact supersymmetry limit,
miss also a solution of so-called $\mu$-problem \cite{mu}.
The $\mu$-term of the order of soft SUSY breaking terms,
in fact should be introduced by hand.

Another theoretical weakness of SUSY GUTs is a lack in the
understanding of flavour. Although GUTs can potentially unify
the Yukawa couplings within one family, the origin of
inter-family hierarchy and weak mixing pattern remains open.
Moreover, in the light families the Yukawa unification
simply contradicts to observed mass
pattern, though the $b-\tau$ Yukawa unification \cite{b-tau}
may constitute a case of partial but significant success.
In order to deal with the flavour problem in GUT frameworks,
some additional ideas (horizontal symmetry, zero textures)
are required \cite{Fritzsch,DHR}.

\vspace{1cm}
{\large \bf 2. GIFT -- Goldstones Instead of Fine Tuning}
\vspace{0.7cm}

An attractive possibility to solve the gauge hierarchy problem
and the related problem of the DT splitting, suggested
in \cite{Inoue,Anselm,BD}, can be simply phrased as follows:
Higgs doublets can appear as Goldstone modes of a
spontaneously broken global symmetry, which is larger than the
local symmetry of the GUT. These doublets, being strictly massless
in the exact SUSY limit, acquire nonzero masses after supersymmetry
breaking and thereby triger the electroweak symmetry breaking.
In ref. \cite{Anselm} this mechanism was elegantly named
as GIFT.

In refs. \cite{Inoue,Anselm} GIFT mechanism was implemented in $SU(5)$
model, by {\em ad hoc} assumption that the Higgs superpotential
has larger global symmetry $SU(6)$. This was done by adding a
singlet superfield $I$ to the minimal Higgs sector of $SU(5)$:
$\Phi(24)+\bar{\phi}(\bar{5})+\phi(5)+I(1)$ is just the
$SU(5)$ decomposition of the $SU(6)$ adjoint representation
$\Sigma(35)$. If one assumes that the Higgs superpotential
has the simple form $W_{Higgs}= M\Sigma^2 + \lambda\Sigma^3$,
then it possesses the $SU(6)$ global symmetry.
The supersymmetric ground state
$\langle \Sigma \rangle = V_G \mbox{diag}(1,1,1,1,-2,-2)$
(one among the other discretely degenerated vacua),
breaking $SU(6)$ down to $SU(4)\times SU(2)\times U(1)$, gives
rise to Goldstone supermultiplets in fragments
$(4,\bar 2)+(\bar{4},2)$. At the same time the gauged part
$SU(5)\subset SU(6)$ breaks down to $SU(3)\times SU(2)\times U(1)$,
so that the fragments $(3,\bar 2)+(\bar{3},2)$ are eaten up by
Higgs mechanism.
The remaining Goldstone fragments $h_1=(1,2)$ and $h_2=(1,\bar 2)$
stay massless until supersymmetry (and thus also
larger global symmetry of Higgs Potential) remains unbroken.
However, the global $SU(6)$ symmetry in the Higgs sector of $SU(5)$
seems rather artificial. In general the Higgs superpotential of the
fields involved should be
\be{Anselm}
W_{Higgs}= \mu^2 I + M\Phi^2 + M' I^2 + m\bar{\phi}\phi +
\lambda\Phi^3 + \lambda'I^3 + \lambda''\Phi^2 I +
f \bar{\phi}\Phi\phi + f'I\bar{\phi}\phi
\ee
while the $SU(6)$ invariance is equivalent to imposing
the following constraints
\be{sixFT}
\mu=0,~~~~M=M'=m/2,~~~~\lambda=-\sqrt{15/8}\lambda'=
\sqrt{10/3}\lambda''=f/3=-\sqrt{5/6}f'
\ee
Without valid dynamical or symmetry reasons these constraints
look as {\em several} unnatural fine tunings instead of
{\em one} tuning (\ref{fine_tuning}) needed in minimal SUSY $SU(5)$.
Thus, if one remains within the SUSY $SU(5)$ frames,
GIFT\footnote{ {\em Goldstones Instead of Fine Tuning} } is
LOST\footnote{ {\em Lots Of Strange Tunings } }.

A much more attractive scenario is that the larger global symmetry
arises in an accidental way \cite{BD}. In other words,
it should be an automatic consequence of the gauge symmetry
and the field content of the model. Obviously, this requires
extension of the $SU(5)$ gauge symmetry, say to $SU(6)$ \cite{BD},
with the anomaly-free fermion sector consisting of chiral superfields
$(\bar{6}+\bar{6}'+15)_i$, where $i=1,2,3$ is a family index.
The Higgs sector
contains supermultiplets $\Sigma$ and $H+\bar{H}$, respectively
in adjoint 35 and fundamental $6+\bar{6}$ representations.
If the Higgs superpotential has a structure
$W=M\Sigma^2 + \lambda \Sigma^3 + S(\bar{H}H - V_H^2)$, where
$S$ is an auxiliary singlet, then it acquires an extra global
symmetry $SU(6)_\Sigma\times U(6)_H$.\footnote{Notice, however, that
the crossing term $\bar{H}\Sigma H$ is put to zero by hand.}
In the exact SUSY limit the vacuum state has continuous degeneration:
$\langle H \rangle = \langle \bar{H} \rangle = V_H\, (1,0,0,0,0,0)$,
$\langle \Sigma \rangle = V_G U^\dagger
\mbox{diag}\,(1,1,1,1,-2,-2)U$, where $U$ is arbitrary $6\times6$
unitary matrix.\footnote{ In fact, $SU(6)_\Sigma\times SU(6)_H$
is not a global symmetry of a whole Higgs Lagrangian, but only
of the  Higgs superpotential. In particular, the Yukawa as well
as the gauge couplings ($D$-terms) do not respect it. However,
in the exact supersymmetry limit (i) it is
effective for the field configurations on the vacuum valley,
where $D=0$, (ii) it cannot be spoiled by the radiative corrections
from the Yukawa interactions, owing to non-renormalization theorem. }
 If the true vacuum state corresponds
to $U=1$ (as it can appear after SUSY breaking), then these VEVs
break the $SU(6)$ gauge symmetry down to the standard
$SU(3)\times SU(2)\times U(1)$ symmetry. At the same time,
the global symmetry $SU(6)_\Sigma\times U(6)_H$ is spontaneously
broken to $[SU(4)\times SU(2)\times U(1)]_\Sigma\times U(5)_H$,
and the corresponding Goldstone modes are presented by
fragments $[(4,\bar 2)+(\bar{4},2)]_\Sigma \to
[\bar T'(3,\bar 2)+T'(\bar{3},2)+ \bar{h}(1,\bar 2) + h(1,2)]_\Sigma$
and $\ov{5}_{\bar{H}} +5_H \to
[\bar{T}(\bar{3},1)+\bar{h}(1,\bar 2)]_{\bar{H}} + [T(3,1)+h(1,2)]_H $.
Clearly, all these
are eaten up by corresponding fragments in vector superfields of the
gauge $SU(6)$ symmetry, except the combinations
$h_1\propto V_H \bar{h}_\Sigma - 3V_G \bar{h}_{\bar{H}}$,
$h_2\propto V_H h_\Sigma - 3V_G h_H$,  which
remain massless and can be be identified
with the MSSM Higgs doublets.

In order to maintain the
gauge coupling unification, we have to assume $V_H\geq V_G$, so
that $SU(6)$ is first broken by $H,\bar{H}$ down to $SU(5)$.
Then, at the scale $V_G\simeq 10^{16}\,$GeV, the VEV of $\Sigma$
breaks $SU(5)$ down to $SU(3)\times SU(2)\times U(1)$.
As far as the superpotential of $\Sigma$ does not feel the local
symmetry breaking by the $H,\bar H$ VEVs and continues to carry the
global $SU(6)_\Sigma$ symmetry, in the limit $V_H\gg V_G$ the
theory automatically reduces to the $SU(5)$
GIFT model of refs. \cite{Inoue,Anselm}.

After the SUSY breaking enters the game (presumably through the
hidden supergravity sector), the Higgs potential, in addition
to the (supersymmetric) squared $F$ and $D$ terms, includes also
the soft SUSY breaking terms \cite{BFS}.
These are given by  $V_{SB}=AmW_3 + BmW_2 + m^2|\phi_k|^2$,
where $\phi_k$ imply all scalar fields involved, $W_{3,2}$
are terms in superpotential respectively trilinear and bilinear
in $\phi_k$, and $A,B,m$ are soft breaking parameters.
Due to these terms the VEV of $\Sigma$ is shifted, as compared
to the one calculated in the exact SUSY limit, by an amount of
$\sim m$. Through the $\Sigma^3$ term in the superpotential,
this shift gives rise to term $\mu h_1 h_2$. Thus, the GIFT
scenario automatically solves the $\mu$-problem: the
(supersymmetric) $\mu$-term for the resulting MSSM in fact
arises as a result of SUSY breaking, with $\mu\sim m$.

The Higgs doublets acquire also the soft SUSY breaking mass
terms, but not all of them immediately. Clearly, $V_{SB}$
also respects the larger global symmetry
$SU(6)_\Sigma\times U(6)_H$, so that one combination
of the scalars $h_1$ and $h_2$, namely $\tilde{h}=h_1+h_2^\ast$
remains massless as a truly Goldstone boson. However, as far as
SUSY breaking relaxes radiative corrections, the latter will
remove the vacuum degeneracy and provide non-zero mass to
$\tilde{h}$ (situation, much similar
to the case of axion). The effects of radiative
corrections, which lift vacuum degeneracy and lead to the
electroweak symmetry breaking, were studied in ref. \cite{BDM}.
It was shown that GIFT scenario does not imply any upper bound
on the top quark mass, in spite of earlier
claims \cite{Anselm,Goto} and it can go up up to its infrared
fixed limit.

In fact, the $SU(6)$ model \cite{BD} is a minimal extension of the
standard $SU(5)$ model.
At the scale $V_H$
the fermion
content is reduced to the minimal fermion content of standard $SU(5)$.
Indeed, the $SU(5)$ decomposition of various supermultiplets reads
\beqn{SU5frag}
& & H=(5+1)_H,~~~  \bar{H}=(\bar{5}+1)_{\bar{H}},~~~
\Sigma=(24+5+\bar{5}+1)_\Sigma  \nonumber \\
& & \bar{6}_i=(\ov{5}+1)_i,~~~~~
\bar{6}_i^\prime=(\ov{5}+1)_i^\prime,~~~  15_i=(10+5)_i
{}~~~~~~ (i=1,2,3)
\eeqn
so that the fermion sector at this scale consists of
six $\ov{5}$-plets, three 5-plets, three 10-plets and six singlets.
According to the survival hypothesis \cite{surv}, after the
breaking $SU(6)\to SU(5)$ the extra fermions $(5+\ov{5}')_i$
become heavy (with masses $\sim V_H$) owing to the Yukawa couplings
$\Gamma_{ij}15_i\bar{H}\bar{6}'_j$, and decouple from
from the light states which remain in $(\ov{5}+10)_i$ and singlets.
Thus, below the scale $V_H$ we are left with minimal $SU(5)$ GUT
with three standard fermion families.

The $SU(6)$ theory \cite{BD} drastically differs from any other GUT
approaches. Usually, in GUTs the Higgs sector consists of
two different sets:  one is for the GUT symmetry breaking
(e.g. 24-plet in $SU(5)$), while another is just for
the electroweak symmetry breaking and fermion mass generation
(like $5+\bar{5}$ in $SU(5)$). In contrast, in the $SU(6)$
theory no special superfields are indroduced for the second function.
The Higgs sector consisting of the 35-plet and $6+\bar{6}$, is a
minimal one needed for the $SU(6)$ breaking down to the
$SU(3)\times SU(2)\times U(1)$. As for the Higgs doublets, they
arise as Goldstone modes, from the $SU(2)\times U(1)$ doublet
fragments in $\Sigma$ and $H,\bar{H}$.

Due to this reason, their couplings with the
fermion fields have some peculiarities, which could provide new
possibilities towards the understanding of flavour. Namely,
if by chance the Yukawa superpotential also respects the
$SU(6)_\Sigma \times U(6)_H$ global symmetry, then the Higgs doublets,
as Goldstone fields, have {\em vanishing} Yukawa couplings with the
fermions that remain massless after the $SU(6)$ symmetry breaking.
In fact, these are the chiral families $(\ov{5}+10)_i$ of ordinary
quarks and leptons. Yukawa couplings
$15\,\bar{H}\,\bar{6}$ respect extra global symmetry
and cannot generate their masses, so that one has to
invoke the higher order operators scaled by inverse powers of some
large mass $M\gg V_H$. These could appear due to nonperturbative
quantum gravity effects, with $M\sim M_{Pl}$.
Alternatively, they can arise by integrating out some heavy states
with masses above the $SU(6)$ breaking scale. Indeed, such states
in vectorlike (real) representations can naturally present in
SUSY (stringy) GUTs. According to survival hypothesis, they
should acquire maximal allowed masses $M$ if there
are no symmetry reasons to keep them light. Then masses of
ordinary light fermions can appear as a result of 'seesaw' mixing
with these heavy states \cite{Frogatt}. In model \cite{BD},
operators relevant for down quarks appear as
$\frac{1}{M}15\Sigma \bar{H}\bar{6}$ while the operators
relevant for upper quarks are $\frac{1}{M^2}15 H \Sigma H 15$.
So, it seems that model leads to unacceptable case  $m_b\gg m_t$.

As it was shown in ref. \cite{BDSBH}, the problem can be resolved
by introducing a fermion $20$-plet, which $SU(5)$ content is
$20=10+\ov{10}$. Since $20$ is a pseudo-real representation
(the tensor product $20\times 20$ contains singlet only in an
antisymmetric combination), the survival hypothesis does not apply
to it.
More generally, if in original theory $20$-plets present in odd
number then {\em one} of them should inevitably stay massless.
Then its Yukawa couplings $g20\Sigma 20$ and $G_i20 H 15_i$
{\em explicitly} violate the global $SU(6)_\Sigma\times U(6)_H$
symmetry. As a result, the {\em only} particle which gets direct
Yukawa coupling with the 'Goldstone' Higgs doublet is an upper quark
contained in $20$, that is top quark.
Other fermions stay massless at this level, and for generating
their masses one has to appeal to higher order operators.
In order to achieve a proper operator structure,
additional symmetries are needed. On the other hand, consistency
of the GIFT scenario also requires some extra symmetry in order
to forbid the crossing term $\bar{H}\Sigma H$ -- otherwise
the Higgs superpotential has no accidental global symmetry.

Below we describe a consistent SUSY $SU(6)$ model with
flavour-blind discrete $Z_3$ symmetry \cite{BB}.
The role of the latter is important:
first, it guarantees that Higgs superpotential has automatic global
$SU(6)_\Sigma \times U(6)_H$ symmetry without putting to zero of some
terms by hand, and second,
it provides proper structure of the higher order operators
generating realistic mass and mixing pattern for {\em all} fermion
families.

\vspace{1cm}
{\large \bf 2. $SU(6)\times Z_3$ model }
\vspace{0.7cm}

Consider the supersymmetric model with $SU(6)$ gauge symmetry, with
the set of chiral superfilds consisting of two sectors:

$(i)$ The `Higgs' sector: vectorlike supermultiplets $\Sigma_1(35)$,
$\Sigma_2(35)$, $H(6)$, $\bar{H}(\bar{6})$ and an auxiliary
singlet $S$.

$(ii)$ The `fermion' sector: chiral, anomaly free supermultiplets
$(\bar{6} + \bar{6}^\prime)_i$, $15_i$ ($i=1,2,3$ is a family index)
and 20.


We introduce also two flavour-blind discrete symmetries.
One is usual matter parity $Z_2$, under which
the fermion superfields change the sign while the Higgs ones stay
invariant. Such a matter parity, equivalent to R parity,
ensures the proton stability.
Another discrete symmetry is $Z_3$ acting in the following way:
\beqn{Z3}
 & &  \Sigma_1 \to \omega\, \Sigma_1, ~~~
\Sigma_2 \to \bar{\omega}\,\Sigma_2, ~~~
H,\bar{H} \to H,\bar{H}, ~~~ S \to S  \nonumber \\
 & & \bar{6}^i_{1,2} \to \omega\, \bar{6}^i_{1,2}, ~~~
15^i \to \bar{\omega}\, 15^i, ~~~ 20 \to \omega\, 20 ~~~~~
(\omega=\mbox{e}^{i\frac{2\pi}{3}})
\eeqn

Let us consider first the Higgs sector.
The most general renormalizable superpotential compatible with the
$SU(6)\times Z_3$ symmetry is
$W_{Higgs} = W_{\Sigma} + W_H + W(S) $,
where $W(S)$ is a polynomial containing linear, quadratic
and cubic terms, and
\be{superpot}
W_\Sigma = M \Sigma_1 \Sigma_2 + \lambda_0 S \Sigma_1 \Sigma_2 +
\lambda_1 \Sigma_1^3 + \lambda_2 \Sigma_2^3\,, ~~~~~
W_H = M^\prime \bar{H}H + \lambda S\bar{H}H
\ee
This superpotential automatically has larger global symmetry
$SU(6)_{\Sigma}\times U(6)_H$, related to independent transformations
of the $\Sigma$ and $H$ fields. In the exact supersymmetry limit,
the condition of vanishing $F$ and $D$ terms allows, among other
discretely (and continuously) degenerated vacua, the VEV
configuration
\beqn{VEVS}
&& \langle\Sigma_{1,2}\rangle = V_{1,2}\,\mbox{diag}(1,1,1,1,-2,-2)
\nonumber \\
&& \langle H\rangle = \langle\bar{H}\rangle = V_H\,(1,0,0,0,0,0)
\eeqn
For a proper parameter range, these configuration
can appear as a true vacuum state afterthat the vacuum
degeneracy is lifted by soft SUSY breaking and subsequent
radiative corrections \cite{BDM}.

The VEVs (\ref{VEVS}) lead to the needed pattern of gauge symmetry
breaking: $H,\bar{H}$ break $SU(6)$ down to $SU(5)$, while
$\Sigma_{1,2}$ break $SU(6)$ down to $SU(4)\times SU(2)\times U(1)$.
Both channels together break the local symmetry down to
$SU(3)\times SU(2)\times U(1)$. At the same time, the global
symmetry $SU(6)_{\Sigma}\times U(6)_H$ is broken down to
$[SU(4)\times SU(2)\times U(1)]_\Sigma \times U(5)_H$.
The Goldstone degrees which survive from being eaten by Higgs
mechanism constitute the couple $h_1+h_2$ of the MSSM Higgs
doublets, which in
terms of the doublet (anti-doublet) fragments in $\Sigma_{1,2}$
and $H,\bar{H}$ are given as
\beqn{Higgs}
h_2= \cos\alpha(\cos\gamma\,h_{\Sigma_1} +
\sin\gamma\,h_{\Sigma_2}) + \sin\alpha\, h_H  \nonumber \\
h_1= \cos\alpha(\cos\gamma\,\bar{h}_{\Sigma_1} +
\sin\gamma\,\bar{h}_{\Sigma_2}) + \sin\alpha\, \bar{h}_{\bar{H}}
\eeqn
where $\tan\gamma=V_2/V_1$ and $\tan\alpha=3V_G/V_H$,
$V_G=(V_1^2+V_2^2)^{1/2}$.
In the following  we adopt the case $V_H\gg V_1\gg V_2$.
Thus, in this case the Higgs doublets dominantly come from $\Sigma_1$,
while in $\Sigma_2$ and $H,\bar{H}$ they are contained with small
weights $\eps_2/\eps_1$ and $3\eps_1$ respectively,
where $\eps_{1,2}=V_{1,2}/V_H$.

\vspace{1cm}
{\large \bf 4. Fermion masses }
\vspace{0.7cm}

The most general Yukawa superpotential compatible with the
$SU(6)\times Z_3$ symmetry has the form
\be{Yukawa}
W_{Yuk} = g\,20 \Sigma_1 20 \,+ \,G\,20 H 15_3\,  + \,
\Gamma_{ij} 15_i \bar{H} \bar{6}^\prime_j\, ~~~~~~i,j=1,2,3
\ee
where all Yukawa coupling constants are assumed to be $O(1)$
(for comparison, we remind that the gauge coupling constant at GUT
scale is $g_{GUT}\simeq 0.7$). Without loss of generality,
one can always redefine the basis in 15-plets so that
only the $15_3$ state couples to 20-plet in (\ref{Yukawa}).
Also, among six $\bar{6}$-plets
one can always choose the basis when only three of them (denoted
in eq. (\ref{Yukawa}) as $\bar{6}^\prime_{1,2,3}$) couple to
$15_{1,2,3}$, while other three states $\bar{6}_{1,2,3}$ have
no Yukawa couplings.

For $V_H \gg V_G$, already at the breaking $SU(6) \to
SU(5)$, the light fermion states are identified from this
superpotential, whereas the extra fermion states become superheavy.
Indeed, the $SU(5)\supset SU(3)\times SU(2)\times U(1)$ decomposition of
the fermion multiplets under consideration reads
\beqn{fragments}
& & 20=10 + \ov{10} = (q+u^c+e^c)_{10}+(Q^c+U+E)_{\ov{10}} \nonumber \\
& & 15_i=(10+5)_i = (q_i+u^c_i+e^c_i)_{10} +
(D_i + L^c_i)_5  \nonumber \\
& & \bar{6}_i=(\bar{5}+1)_i = (d^c_i + l_i)_{\bar{5}} + N_i \nonumber \\
& & \bar{6}_i^\prime=(\bar{5}+1)_i^\prime =
(D^c_i +L_i)_{\bar{5}^\prime} + N_i^\prime
\eeqn
According to eq. (\ref{Yukawa}), the extra fermion pieces with
non-standard $SU(5)$ content, namely $\ov{10}$ and $5_{1,2,3}$,
form massive particles being coupled with $10_3$ and $\ov{5}'_{1,2,3}$
\be{heavymass}
G\,V_H\,\ov{10}\,10_3\, +
\,\Gamma_{ij}V_H\,5_i\,\bar{5}_j^\prime\, +
\,g\, V_1 \,(U\,u^c - 2 E\,e^c)
\ee
and thereby decouple from the light states which remain in
$\bar{5}_{1,2,3}$, $10_{1,2}$ and 10 (if neglect the small,
$\sim \eps_1$ mixing
between the $u^c - u^c_3$ and $e^c - e^c_3$ states).
On the other hand, since the couplings of 20-plet explicitly
violate the global $SU(6)_\Sigma\times U(6)_H$ symmetry,
the Higgs doublet $h_2$ has {\em non-vanishing} couplings with
up-type quarks from 20- and $15_3$-plets.
Indeed, it follows from eq. (\ref{Yukawa}) that {\em only} Yukawa
coupling relevant for light states is contained in
$g 20\Sigma_1 20 \to g 10\,5_{\Sigma_1} 10$. Therefore, only one
up-type quark (to be identified with top quark), dominantly
contained in 20-plet, has a direct Yukawa coupling
$~g\, q u^c \,h_2,~$
so that its mass has to be in the 100 GeV range.
Other fermions stay massless at
this level, unless we invoke the higher order operators
explicitly violating the accidental global symmetry.

Higher order operators scaled by inverse powers of some
large mass $M\gg V_H$ could appear due to quantum gravity effects,
with $M\sim M_{Pl}$. Alternatively, they can arise by integrating
out some heavy states with masses $M\gg V_H$.
In the Sect.~5  we adopt the second viewpoint, namely that
these operators appear from the exchange of some heavy fermion
superfields \cite{Frogatt}. The reason is twofold: first,
as we see shortly, the fermion mass pattern favours the scale
$M\sim 10^{18}\,$GeV (string scale?) rather than $M_{Pl}$, and second,
the mechanism of heavy fermion exchange is rather instructive
for obtaining the realistic fermion mass pattern.

Before addressing the concrete scheme of heavy fermion
exchanges, let us start with the general operator analysis.
$Z_3$ symmetry forbids any `Yukawa' operator in the superpotential
at $1/M$ order.\footnote{ Operators involving an odd number of
fermion superfields are forbidden by matter parity.}~
However, operators at the next ($1/M^2$) order are allowed and they
are the following:
\be{dim4}
{\cal A} = \frac{a}{M^2} 20 \bar{H} \Sigma_1 \bar{H} \bar{6}_3\,,~~~~
{\cal B} = \frac{ b_{ij} }{M^2} 15_i H \Sigma_2 H 15_j\,,~~~~
{\cal C} = \frac{c_{ik} }{M^2} 15_i(\Sigma_1\Sigma_2\bar{H})\bar{6}_k
\ee
where $a,b,c$ are $O(1)$ `Yukawa' constants.
Analogous operators involving heavy $\bar{6}^\prime_i$ states are
irrelevant for the light fermion masses.
According to eq. (\ref{heavymass}) the state
$10_3\subset 15_3$ is also heavy, and it is decoupled from the light
particle spectrum. Therefore, operators (\ref{dim4}) are relevant
only for $10\subset 20$,
$10_i\subset 15_i$ ($i=1,2$) and $\bar{5}_k\subset \bar{6}_k$
($k=1,2,3$) states. Without loss of generality, we redefine the basis
of $\bar{6}$-plets so that only the $\bar{6}_3$ state couples to
20-plet in eq. (\ref{dim4}).
It is worth to note that in fact ${\cal C}$ can contain two relevant
combinations with different convolutions of the $SU(6)$ indices
indicated by brackets in an obvious way:
\be{C12}
C_1\!=\!15 (\Sigma_1\Sigma_2 \bar{H}) \bar{6},~~~
C_2\!=\!15 (\Sigma_1 \bar{H}) (\Sigma_2 \bar{6}),~~~
C_3\!=\!(15\bar{H})(\Sigma_1\Sigma_2 \bar{6}),~~~
C_4\!=\!(15\bar{H}\bar{6})(\Sigma_1\Sigma_2)
\ee
$C_1$ and $C_2$ provide different Clebsch coefficients for the down
quark and charged lepton mass terms, while $C_3$ and $C_4$ are
irrelevant for the mass generation and they lead only to some minor
rotation of the heavy fermion states.

Let us analyse now the impact of these operators on the fermion masses.
Obviously, the operator ${\cal A}$ is responsible for
the $b$ quark and $\tau$ lepton masses, and at the MSSM level
it reduces to Yukawa couplings
$a\eps_H^2 \,(q d^c_3  + e^c l_3)\,h_1$, where $\eps_H=V_H/M$.
Hence, though $b$ and $\tau$ belong to the same
family as $t$ (namely, to 20-plet), their Yukawa couplings are
substantially (by factor $\sim \eps_H^2$) smaller than
$\lambda_t\approx g$.
Moreover, we automatically have almost precise $b-\tau$ Yukawa
unification at the GUT scale:
\be{bottom}
\lambda_b = a\eps_H^2 = \lambda_\tau [1+(\eps_1 g/G)^2]
\ee
where the $\sim \eps_1^2$ correction is due to mixing between
the $e^c$ and $e^c_3$ states (see eq. (\ref{heavymass})).

As far as the fermions of the third family are already defined
as the states belonging to
$10\subset 20$ and $\bar{5}_3\subset \bar{6}_3$,
the operators ${\cal B}$ and ${\cal C}$ generate mass terms
for the fermions of the first two families, which
in general case are expected to be of the same order.
In order to achieve mass splitting between the second and first
families, one can assume that the `Yukawa' matrices
$b_{ij}$ and $c_{ik}$ are {\em rank-1} matrices, so that each of
operators ${\cal B}$ and ${\cal C}$ can provide only one non-zero
mass eigenvalue (i.e. $c$ and $s$ quark masses).
Then, without loss of generality, we can redefine the basis of
$15_{1,2}$ and $\bar{6}_{1,2}$ states so that these matrices have
the form
\beqn{product}
&& ~~~~~~~~
b_{ij}=(0,\beta)^T\bullet (0,\beta)= \mat{0}{0}{0}{b} \nonumber \\
&&
c_{ik}=(\gamma_1,\gamma_2)^T \bullet (0,\delta_2,\delta_3)=
\left(\begin{array}{ccc}{0}&{c_2\sin\theta}&{c_3\sin\theta}\\
{0}&{c_2\cos\theta}&{c_3\cos\theta}\end{array}\right)
\eeqn
where $\tan\theta=\gamma_1/\gamma_2$. Hence, in this basis
only $b_{22}=b$ component of the symmetric matrix $b_{ij}$ is nonzero,
and $c$ quark should be identified as an up-quark state from
$q_2,u^c_2 \subset 10_2 \subset 15_2$. Then $s$ quark state is the
down quark state in $q'_2\subset 10'_2\subset 15'_2$ and
$d^c_2 \subset \bar{5}_2 \subset \bar{6}_2$, where
$15'_2=\sin\theta\cdot 15_1 + \cos\theta\cdot 15_2$ is an effective
combination of the $15_{1,2}$ states which couples $\bar{6}_2$ and
$\bar{6}_3$ states (it is not difficult to recognize
that in fact $\theta$ is the Cabibbo angle, which in general tends
to be $O(1)$). Clearly, $\mu$-lepton is also contained in $15'_2+
\bar{6}_2$. Thus, operators ${\cal B}$ and ${\cal C}$ reduce to
MSSM Yukawa couplings for the second family quarks and leptons
\be{charm}
b(\eps_2/\eps_1)\eps_H^2 \,q_2 u^c_2 \,h_2,~~~~
c_{2,3}\eps_2 \eps_H^2\, (q'_2 d^c_{2,3} +
K e'^c_2 l_{2,3})\,h_1
\ee
where $K$ is the Clebsch coefficient
depending on weights of operators (\ref{C12})
entering ${\cal C}$ (for example, $K=1$ if ${\cal C}\propto C_1$
and $K=-2$ if ${\cal C}\propto C_2$).

For the first family fermion masses one can
appeal to $1/M^3$ operators which can be presented as following:
\be{dim5}
{\cal D}=\frac{d_{ik}}{M^3} 15'_i \Sigma_{1}^3 \bar{H} \bar{6}_k\,,
{}~~~~~{\cal E} = \frac{e_{ij}}{M^3} 15_i H \Sigma_1^2 H 15_j
\ee
where $15'_1$ is a state orthogonal to $15'_2$.
As in the case of operator ${\cal C}$, these operators
can have different convolutions of the $SU(6)$ indices.
For ${\cal D}$ the relevant combinations, giving different
relative Clebsch factors $P$ for the $d$ quark and electron masses, are
\be{Dvariants}
D_1= 15(\Sigma_1^3\bar{H})\ov{6},~~~
D_2= 15(\Sigma_1^2\bar{H})(\Sigma_1\ov{6}),~~~
D_3= 15(\Sigma_1\bar{H})(\Sigma_1^2\ov{6}),~~~
D_4= 15(\Sigma_1\bar{H})\ov{6}(\Sigma_1^2)
\ee
Operator ${\cal D}$, with arbitrary couplings $d_{ij}$
of the order of 1, will provide $d$-quark and electron masses
in the correct range. As for the operator ${\cal E}$, for
$e_{11}\sim 1$, it leads to somewhat large value of $u$-quark mass.
Therefore,
it is more suggestive to think that only $e_{12}$ couplings are
non-zero, while $e_{11}=0$. As we show in Sect.~5,
this is really the case in the context of heavy fermion exchange
model.

Keeping only the leading contributions to each component,
for the quark and lepton Yukawa couplings at the GUT scale we obtain
\begin{eqnsystem}{sys:ude}
&&\bordermatrix{& u^c_1 & u^c_2 & u^c \cr
q_1 & 0 & e_{12}\eps_1\eps_H^3 & 0 \cr
q_2 & e_{12}\eps_1\eps_H^3 & b (\eps_2/\eps_1)\eps_H^2 & 0 \cr
q   & 0 & 0 & g \cr} \cdot h_2 \\
&&
\bordermatrix{& d^c_1&d^c_2 & d^c_3 \cr
q'_1  & d_{11} \eps_1^2\eps_H^3 & d_{12} \eps_1^2\eps_H^3 &
        d_{13} \eps_1^2\eps_H^3   \cr
q'_2  & d_{21} \eps_1^2\eps_H^3 & c_2\eps_2\eps_H^2 &
        c_3\eps_2\eps_H^2 \cr
q     & 0 & 0 & a\eps_H^2} \cdot h_1 \\
&&
\bordermatrix{& l_1 & l_2 & l_3 \cr
e'^c_1 & Pd_{11} \eps_1^2\eps_H^3 & Pd_{12} \eps_1^2\eps_H^3 &
        Pd_{13} \eps_1^2\eps_H^3   \cr
e'^c_2 & Pd_{21} \eps_1^2\eps_H^3 & Kc_2\eps_2\eps_H^2 &
        Kc_3\eps_2\eps_H^2 \cr
e^c   & 0 & 0 & a\eps_H^2} \cdot h_1
\end{eqnsystem}
Thus, the Yukawa coupling eigenvalues at the GUT scale are
$\lambda_t=g\sim 1$ and
\beqn{Yukawas}
& \lambda_b=\lambda_\tau=a\eps_H^2 ~~~& \Rightarrow ~~~
\lambda_b/\lambda_t\sim \eps_H^2  \nonumber \\
& \lambda_c=b (\eps_2/\eps_1)\eps_H^2 ~~~& \Rightarrow ~~~
\lambda_c/\lambda_b\sim \frac{\eps_2}{\eps_1}  \nonumber \\
& \lambda_s=\lambda_\mu/K =c_2 \eps_2\eps_H^2 ~~~& \Rightarrow ~~~
\lambda_s/\lambda_b\sim \eps_2   \nonumber \\
& \lambda_d=\lambda_e/P = d_{11}\eps_1^2\eps_H^3 ~~~& \Rightarrow ~~~
\lambda_d/\lambda_s\sim
\eps_2\left(\frac{\eps_1}{\eps_2}\sqrt{\eps_H}\right)^2 \nonumber \\
& \lambda_u=\frac{e_{12}^2}{b}(\eps_1^3/\eps_2)\eps_H^4 ~~~&
\Rightarrow ~~~ \lambda_u/\lambda_d\sim \frac{\eps_2}{\eps_1}
\left(\frac{\eps_1}{\eps_2}\sqrt{\eps_H}\right)^2
\eeqn
In order to connect these Yukawa constants to the
physical masses of the quarks and leptons, the renormalization group
(RG) running has to be considered \cite{DHR,Barger}. We have:
\beqn{masses}
& &
m_t = \lambda_t A_u \eta_t y^6 v\sin\beta, ~~~~
m_b = \lambda_b A_d \eta_b y v\cos\beta, ~~~~
m_\tau = \lambda_\tau A_e \eta_\tau v\cos\beta   \nonumber \\
& &
m_c = \lambda_c A_u \eta_c y^3 v\sin\beta, ~~~~
m_s = \lambda_s A_d \eta_d v\cos\beta, ~~~~
m_\mu= \lambda_\mu A_e \eta_l v\cos\beta  \nonumber \\
& &
m_u = \lambda_u A_u \eta_u y^3 v\sin\beta, ~~~~
m_d = \lambda_d A_d \eta_d v\cos\beta, ~~~~
m_e = \lambda_e A_e \eta_l v\cos\beta
\eeqn
where $v=174\,$GeV and $\tan\beta=v_2/v_1$ is a famous ratio of the
$h_2$ and $h_1$ VEVs. The factors $A_f$ account for the running
induced by gauge couplings from the GUT scale $M_G$
to the SUSY breaking scale $M_S$ (for the definiteness we take
$M_S\simeq m_t$), and $y$ includes the running induced by the
top quark Yukawa coupling:
\be{y}
y=\exp\left[-\frac{1}{16\pi^2}\int_{\ln M_S}^{\ln M_G}
\lambda_t^2(\mu)\mbox{d}(\ln \mu) \right]
\ee
The factors $\eta_f$ encapsulate the running from $M_S$ down
to $m_f$ for heavy quarks $f=t,b,c$ or $\mu=1\,$GeV for light quarks
$f=u,d,s$. Taking all these into the account,
we see that quolitatively correct description of {\em all} fermion
masses can be achieved with
$\eps_H=V_H/M \sim 0.1$, $\eps_1=V_1/V_H \sim 0.1$ and
$\eps_2/\eps_1=V_2/V_1 \sim 0.3$
(so that $\frac{\eps_1}{\eps_2}\sqrt{\eps_H}\sim 1$),
provided that $\tan\beta=1-1.5$. Interestingly, this region of
$\tan\beta$ is favoured by the electroweak symmetry radiative breaking
picture  in the presence of $b-\tau$ Yukawa unification (see eq.
(\ref{bottom})). In addition, $b-\tau$ unification and small
$\tan\beta$ require substantially large $\lambda_t$, actually close
to its infrared fixed point \cite{IRfixed}, which implies for the
physical top mass $M_t\approx \sin\beta (190-210)\,$GeV.
As far as  the scale
$V_1\simeq 10^{16}\,$GeV is fixed by the $SU(5)$ unification
of gauge couplings, these relations in turn imply that
$V_H\sim 10^{17}\,$GeV and $M \sim 10^{18}\,$GeV.

Obtained mass matrices give rise to a quolitatively correct picture
of quark mixing. In particular, one obtains the CKM matrix at the
unification scale as
\be{CKM}
V_{\rm CKM}\approx
\matr{ 1 }{ s_{12} }{ s_{12}s_{13} - s_{13}e^{-i\delta} }
{ -s_{12} }{ 1 }{ s_{23} + s_{12}s_{13}e^{-i\delta} }
{ s_{13}e^{i\delta} }{ -s_{23} }{ 1 }
\ee
where $\delta$ is a CP-phase and
\beqn{angles}
& s_{12}\approx \gamma_1/\gamma_2 \sim 1 ~~~~
& (\vert V_{us}\vert=0.220\pm0.002) \nonumber \\
& s_{23}\approx c_3\lambda_s/c_2\lambda_b \sim \eps_2 ~~~~
& (\vert V_{cb}\vert=0.04\pm0.01) \nonumber \\
& s_{13}\approx d_{13}\lambda_d/d_{11}\lambda_b \sim \eps_2^2 ~~~~
& (\vert V_{td}\vert = 0.03-0.1)
\eeqn
for comparison, in the brackets the 'experimental'
values of mixing angles are shown. The questions of the neutrino
mass pattern and the proton decay features due to $d=5$ Higgsino
mediated operators are considered in refs. \cite{BDSBH,BB}.


\vspace{1cm}
{\large \bf  5. Yukawa couplings generated by heavy particle exchanges}
\vspace{7mm}

{}From the previous section, we are left with two problems:
the difficulty in splitting the masses of the first
two families (in Sect.~4 the form (\ref{product}) for the
coupling constants in operators ${\cal B}$ and ${\cal C}$ was
assumed by hand), and
the need to suppress the coupling $e_{11}$ in operator
${\cal E}$, which leads to unacceptably large $u$ quark mass.

Here we show how both problems can be solved, still without
appealing to any flavour symmetry, by assuming that all higher order
operators are generated by the exchanges of some heavy
superfields with $\sim M$ masses. As we see below, this mechanism
provides also specific predictions for the Clebsch coefficients
$K$ and $P$ distinguishing down quark and charged lepton masses.

Let us introduce the set of heavy vectorlike fermions (in the
following referred as $F$-fermions) with masses $O(M)$ and
transformation properties under $SU(6)\times Z_3$ given in Table 1.
Certainly, we prescribe negative matter parity to all of them.

The operators ${\cal A}$, ${\cal B}$, ${\cal C}$ obtained by the
proper exchange chains are shown in Fig.~1.
We see that these operators generate
only the third and second family masses. Indeed,
coupling $15_F\Sigma_1 \ov{15}_F^1$ defines $\ov{15}_F^1$ state
while the corresponding $15_F^1$ state defines $\ov{6}_3$
through the coupling $15_F^1\bar{H}\ov{6}_3$. The operator
${\cal A}$ is unambiguosly built in this way. On the other
hand, coupling $15_2 H 20_F$ defines $15_2$ state, so that the
operator ${\cal B}$ contributes only the $c$ quark mass.
The coupling
$(\gamma_1 15_1 + \gamma_2 15_2)\Sigma_1 \ov{15}'_F$ defines
$15'_2$ state, which in general does not coincide with $15_2$,
and the coupling $15_F^2\bar{H}\bar{6}_2$ defines $\ov{6}_2$ state.
Therefore, the operator ${\cal C}$ providing only the $s$ quark and
$\mu$-lepton masses, in general implies the large Cabibbo angle,
$\tan\theta=\gamma_1/\gamma_2$. In addition, the operator ${\cal C}$
derived in this way, acts as combination ${\cal C} \propto
C_1+2C_2$ of operators (\ref{C12}), which leads
to specific Clebsch coefficient $K=-5$ in eq. (\ref{charm}).

Exchanges generating operators ${\cal D, E}$ are shown in Fig.~2.
In reproducing these operators, we have taken into account
the following restriction: ${\cal D}$ built by $F$-fermion exchange
should be irreducible to lower ($1/M^2$) order operator, in order
to guarantee the mass hierarchy between first and second families.
In other words, the exchange chain should not allow to replace
$\Sigma_1\Sigma_1$ by $\Sigma_2$. This condition requires
large representations of $SU(6)$ involved into the exchange. Then the
operator ${\cal D}$ built as shown in Fig.~1D acts in combination
${\cal D} \propto D_1+D_3-D_4$, which gives relative Clebsch
coefficient $P=5/8$ for the $d$ quark and electron masses.
On the other hand, the operator ${\cal E}$ built as in Fig.~2E,
can only mix $15_1$ state containing $u$ quark, with $15_2$
state containing $c$ quark, but cannot provide direct mass
term for the former.

\begin{table}[t]
\begin{center}\begin{tabular}{c|c|c|l}
$Z_3$: & Higgs & fermions & ~~~~ $F$-fermions \\ \hline
$\omega$ & $\Sigma_1$ & $\bar{6}_i$, $\bar{6}'_i$, $20$ &
$15_F$,~ $\overline{15}_F^{1,2}$,~ $20_F$,~ $84_F$ \\ \hline
$\bar{\omega}$ & $\Sigma_2$ & $15_i$ & $\overline{15}_F$,~
$15_F^{1,2}$,~ $\overline{20}_F$,~ $\overline{84}_F$ \\ \hline
{\em inv.} & $S$, $H$, $\bar{H}$ & -- & $\overline{15}'_F$,~
$15'_F$,~~ $20_F^{1,2}$,
$\overline{105}_F$,~ $105_F$,~ $\overline{210}_F$,~ $210_F$
\end{tabular}

\caption{$Z_3$-transformations of various supermultiplets.}
\end{center}\end{table}

As a result, the higher order operators obtained by the exchange of
$F$-fermions given in Table~1, consistently reproduce the
mass matrix ansatz given in Sect.~4. Moreover, specific Clebsch
coefficients are obtained, leading to relations
$\lambda_d=\frac{1}{5}\lambda_\mu$ and
$\lambda_d=\frac{8}{5}\lambda_e$ (small ($\sim \eps_1$)
corrections to these can arise from the interference of the operators
${\cal C}$ and ${\cal D}$). According to eqs. (\ref{masses}), these
relations imply
\be{d/s}
\frac{m_d}{m_s}\simeq 8\,\frac{m_e}{m_\mu}\approx \frac{1}{25}
\ee
In addition, by taking into account the uncertainties of
renormalization factors (\ref{masses}), mainly due to uncertainty
in $\alpha_3(M_Z)$,
for the quark running masses at $\mu=1\,$GeV we obtain
\be{s_mass}
m_s =90-150\,\mbox{MeV},~~~~~
m_d =4-7\,\mbox{MeV}
\ee
in agreement with the experimental values.

Let us conclude with following remark. As we have seen, the
fermion mass pattern requires that scales $M$, $V_H$ and $V_G$
are related as $V_G/V_H \sim V_H/M \sim 0.1$. The superpotential
(\ref{superpot}) includes mass parameters, which are not related
to $M$. Therefore, it cannot explain why the scales should be
arranged in this way. Bearing in mind the possibility that
considered $SU(6)$ theory could be a stringy
SUSY GUT, one can assume that the superfields $H,\bar H$ and
$\Sigma_{1,2}$ are zero modes, and their superpotential has
the form not containing mass terms:
\be{new_W}
W=S[\bar H H -(\eps_H M)^2] + \lambda_1\Sigma_1^3 +
\lambda_2\Sigma_2^3 + \frac{(\bar H H)}{M}\,(\Sigma_1\Sigma_2)
\ee
The last term can be effectively obtained by exchange of
the singlet superfield $Z$ with a large mass term $MZ^2$,
as shown in Fig.~3. Then the relation $V_G/V_H \sim V_H/M=\eps_H$
follows naturally. Certainly, the origin of small linear
term ($\eps_H\sim 0.1$) in (\ref{new_W}) remains unclear.
It may arise due to some hidden sector outside the GUT.

Non-perturbative effects in principle could induce the higher
order operators scaled by inverse powers of the Planck mass.
If all such operators unavoidably occur, this would
spoil the GIFT picture. For example, already the operator
$\frac{1}{M_{Pl}} (\bar H \Sigma_1)(\Sigma_2 H)$ would provide
an unacceptably large ($\sim M_G^2/M_{Pl}$) mass to the Higgs
doublets. One may hope,
however, that not all possible structures appear in higher
order terms. Alternatively, one could try to suppress dangerous
high order operators by symmetry reasons, in order to achieve
a consistent 'all order' solution \cite{BCL}.

\vspace{1.3cm}
{\large \bf Acknowledgements}
\vspace{0.7cm}

I thank Riccardo Barbieri and Gia Dvali for
fruitfull discussions and collaboration on this subject,
and Dr. Ursula Miscili for encouragement.




\newpage
\baselineskip=12pt

\begin{figure}\setlength{\unitlength}{1.3cm}
\begin{center}\begin{picture}(6,5.7)(0,-5)
\put(-1,-0){\makebox(0,0){ {\cal A:} }}
\put(-1,-2){\makebox(0,0){ {\cal B:} }}
\put(-1,-4){\makebox(0,0){ {\cal C:} }}
\put(0,0){\line(1,0){6}}
\multiput(1,0)(2,0){3}{\line(0,-1){1}}
\put(0,0.1){\makebox(0,0)[bl]{20}}
\put(2,0.1){\makebox(0,0)[b]{$\overline{15}_{\rm F}~~~15_{\rm F}$}}
\put(4,0.1){\makebox(0,0)[b]{$\overline{15}^1_{\rm F}~~~15^1_{\rm F}$}}
\put(6,0.1){\makebox(0,0)[br]{$\bar{6}_3$}}
\put(1.1,-1){\makebox(0,0)[bl]{$\bar{H}$}}
\put(3.1,-1){\makebox(0,0)[bl]{$\Sigma_1$}}
\put(5.1,-1){\makebox(0,0)[bl]{$\bar{H}$}}
\put(2,0){\makebox(0,0){$\times$}}
\put(4,0){\makebox(0,0){$\times$}}
%
%
\put(0,-2){\line(1,0){6}}
\multiput(1,-2)(2,0){3}{\line(0,-1){1}}
\put(0,-1.9){\makebox(0,0)[bl]{$15_2$}}
\put(2,-1.9){\makebox(0,0)[b]{$20_{\rm F}~~~\overline{20}_{\rm F}$}}
\put(4,-1.9){\makebox(0,0)[b]{$\overline{20}_{\rm F}~~~20_{\rm F}$}}
\put(6,-1.9){\makebox(0,0)[br]{$15_2$} }
\put(1.1,-3){\makebox(0,0)[bl]{$H$} }
\put(3.1,-3){\makebox(0,0)[bl]{$\Sigma_2$}}
\put(5.1,-3){\makebox(0,0)[bl]{$H$}}
\put(2,-2){\makebox(0,0){$\times$}}
\put(4,-2){\makebox(0,0){$\times$}}
\put(0,-4){\line(1,0){6}}
\multiput(1,-4)(2,0){3}{\line(0,-1){1}}
\put(0,-3.9){\makebox(0,0)[bl]{$15'_2$} }
\put(2,-3.9){\makebox(0,0)[b]{$\overline{15}'_{\rm F}~~~15'_{\rm F}$}}
\put(4,-3.9){\makebox(0,0)[b]{$\overline{15}^2_{\rm F}~~~15^2_{\rm F}$}}
\put(6,-3.9){\makebox(0,0)[br]{$\bar{6}_{2,3}$}}
\put(1.1,-5){\makebox(0,0)[bl]{$\Sigma_1$}}
\put(3.1,-5){\makebox(0,0)[bl]{$\Sigma_2$}}
\put(5.1,-5){\makebox(0,0)[bl]{$\bar{H}$}}
\put(2,-4){\makebox(0,0){$\times$}}
\put(4,-4){\makebox(0,0){$\times$}}
\end{picture}
\caption{diagrams giving rise\label{Diagrams}
to the operators ${\cal A,~ B,~ C}$ respectively.}
\end{center}\end{figure}
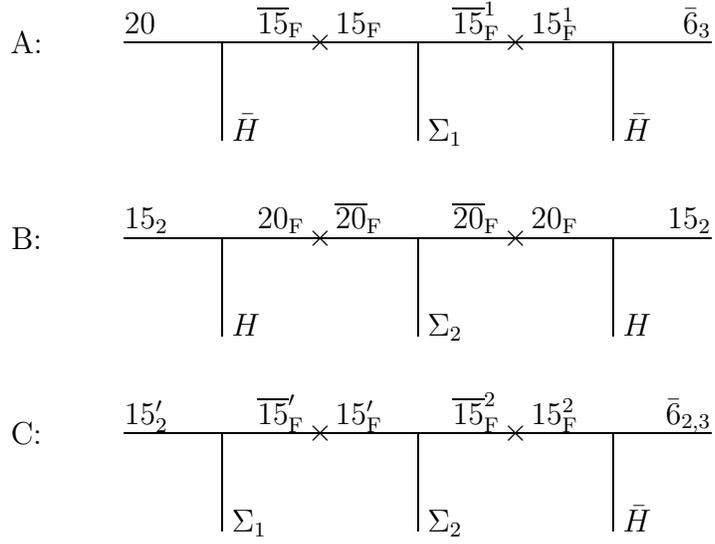



\begin{figure}\setlength{\unitlength}{1.3cm}
\begin{center}\begin{picture}(6,4)(0,-3.3)
\put(-1,-0){\makebox(0,0){ {\cal D:} }}
\put(-1,-2){\makebox(0,0){ {\cal E:} }}
%
\put(0,0){\line(1,0){8}}
\multiput(1,0)(2,0){4}{\line(0,-1){1}}
\put(0,0.1){\makebox(0,0)[bl]{$15_{i}$}}
\put(2,0.1){\makebox(0,0)[b]{$\overline{105}_{\rm F}~~105_{\rm F}$}}
\put(4,0.1){\makebox(0,0)[b]{$\overline{210}_{\rm F}~~210_{\rm F}$}}
\put(6,0.1){\makebox(0,0)[b]{$\overline{84}_{\rm F}~~~84_{\rm F}$}}
\put(8,0.1){\makebox(0,0)[br]{$\bar{6}_{k}$}}
\put(1.1,-1){\makebox(0,0)[bl]{$\Sigma_1$} }
\put(3.1,-1){\makebox(0,0)[bl]{$\bar{H}$} }
\put(5.1,-1){\makebox(0,0)[bl]{$\Sigma_1$} }
\put(7.1,-1){\makebox(0,0)[bl]{$\Sigma_1$} }
\put(2,0){\makebox(0,0){$\times$}}
\put(4,0){\makebox(0,0){$\times$}}
\put(6,0){\makebox(0,0){$\times$}}
\put(0,-2){\line(1,0){8}}
\multiput(1,-2)(2,0){4}{\line(0,-1){1}}
\put(0,-1.9){\makebox(0,0)[bl]{$15_{i}$}}
\put(2,-1.9){\makebox(0,0)[b]{$\overline{105}_{\rm F}~~105_{\rm F}$}}
\put(4,-1.9){\makebox(0,0)[b]{$20^1_{\rm F}~~~20^2_{\rm F}$}}
\put(6,-1.9){\makebox(0,0)[b]{$\overline{20}_{\rm F}~~~20_{\rm F}$}}
\put(8,-1.9){\makebox(0,0)[br]{$15_2$}}
\put(1.1,-3){\makebox(0,0)[bl]{$\Sigma_1$} }
\put(3.1,-3){\makebox(0,0)[bl]{$H$} }
\put(5.1,-3){\makebox(0,0)[bl]{$\Sigma_1$} }
\put(7.1,-3){\makebox(0,0)[bl]{$H$} }
\put(2,-2){\makebox(0,0){$\times$}}
\put(4,-2){\makebox(0,0){$\times$}}
\put(6,-2){\makebox(0,0){$\times$}}
%
%
\end{picture}
\caption{diagrams giving rise\label{Diagrams2}
to the operators ${\cal D}$ and ${\cal E}$ respectively.}
\end{center}\end{figure}
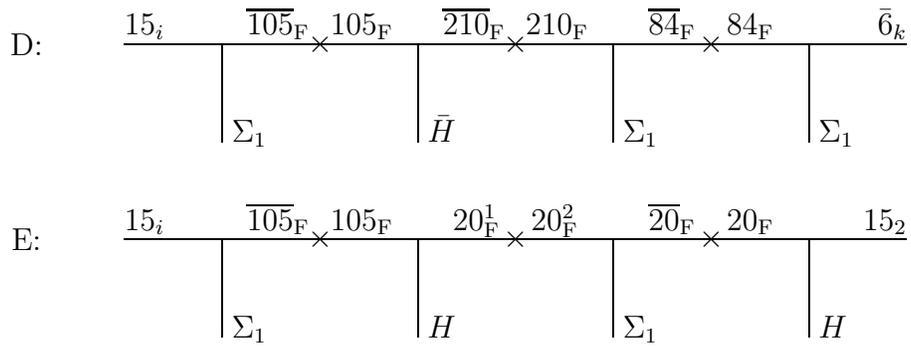



\begin{figure}\setlength{\unitlength}{1.5mm}
\begin{center}\begin{picture}(40,20)
\put(0,0){\vector(1,1){5}}   \put(5,5){\line(1,1){5}}
\put(0,20){\vector(1,-1){5}}\put(5,15){\line(1,-1){5}}
\put(10,10){\line(1,0){20}}
\put(40,20){\vector(-1,-1){5}} \put(35,15){\line(-1,-1){5}}
\put(40,0){\vector(-1,1){5}}\put(35,5){\line(-1,1){5}}
\put(15,10){\vector(-1,0){0}}\put(15,11){\makebox(0,0)[b]{$Z$}}
\put(25,10){\vector(1,0){0}}\put(25,11){\makebox(0,0)[b]{$Z$}}
\put(6,15){\makebox(0,0)[bl]{$\Sigma_1$}}
\put(6,5){\makebox(0,0)[tl]{$\Sigma_2$}}
\put(34,15){\makebox(0,0)[br]{$H$}}
\put(34,5){\makebox(0,0)[tr]{$\bar H$}}
\put(20,10){\makebox(0,0){$\times$}}
\end{picture}
\caption{Diagram generating the operator
$\frac{1}{M}(\bar H H)(\Sigma_1\Sigma_2)$.\label{pDecDiagram}}
\end{center}\end{figure}
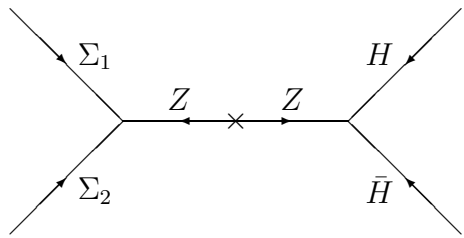


\end{document}